\documentstyle[12pt]{article}
\begin{document}
\title{
\begin{flushright}
{\small SMI-20/96}
\end{flushright}
\vspace{0.5cm}
Improved Algorithm for Bosonized Fermion Determinant.
\author{ A.A.Slavnov \thanks{E-mail:$~~$ slavnov@class.mian.su} \\ Steklov
Mathematical Institute, Russian Academy of Sciences,\\ Vavilov st.42,
GSP-1,117966, Moscow, Russia }} \maketitle

\begin{abstract}

Following the line of ref. \cite{AS} we propose an improved algorithm
which allows to calculate a $D$-dimensional fermion determinant
integrating the exponent of $D+1$-dimensional Hermitean bosonic effective 
action. For a finite extra dimension the corrections decrease exponentially.

\end{abstract}

\section {Introduction}

Recently we proposed an algorithm which allows to calculate $D$-dimensional
fermion determinants by integrating a $D+1$-dimensional bosonic
effective action \cite{AS}. The following representation for the square
of the covariant Dirac operator was obtained:
 \begin{equation} \det( \hat{D}+m)^2 =
\label{1} \end{equation} $$ = \lim_{L \rightarrow \infty, b \rightarrow
0} \int \exp \{a^4b \sum_{n=-N+1}^N \sum_x [b^{-2}( \phi^*_{n+1}(x)
\phi_n(x)+ \phi^*_n(x) \phi_{n+1}(x)-2 \phi^*_n \phi_n) + $$ $$ +i[
\phi^{*}_{n+1}(x) \gamma_5( \hat{D}+m) \phi_n(x)- \phi^{*}_n(x) \gamma_5
 (\hat{D}+m) \phi_{n+1}(x)][b]^{-1}+ $$ $$ +\frac{1}{2}
\phi^{*}_n(D^2-m^2) \phi_n +\frac{i}{ \sqrt{L}}( \phi^*_n(x) \chi(x)+
\chi^*(x) \phi_n(x))] \}d \phi^*_nd \phi_nd \chi^*d \chi .  $$
Here $ \phi_n$ are bosonic fields defined on a 4+1 dimensional lattice
which have the same spinorial and internal structure as the Dirac fields
$ \psi$. 
 The fifth component $t$ to be
defined on the one dimensional lattice of the length $L$ with the lattice
spacing $b$:  \begin{equation} L=2Nb, \quad -N<n \leq N \label{2}
\end{equation}    The free
 boundary conditions in $t$ are assumed:
\begin{equation}
 \phi_n=0, \quad n \leq -N, \quad n>N  \label{3}
\end{equation}
Lattice covariant derivative is denoted by $D_{ \mu}$
 \begin{equation} D_{\mu} \psi(x)=
\frac{1}{a}[U_{\mu}(x) \psi(x+a_{\mu})- \psi(x)] \label{4}
\end{equation}
\begin{equation}
\hat{D}=1/2 \gamma_{ \mu}(D_{ \mu}^*-D_{ \mu}) \label{5}
\end{equation}
(For simplicity we consider naive fermions, but all the construction is
extended in a straightforward way to Wilson fermions).

For finite $L$ and $b$ the eq.(\ref{1}) should be corrected by the terms
\begin{equation}
O(b^2D^2)- \frac{4}{L^2D^2} \sin^2( \frac{DL}{2}) \label{6}
\end{equation}
where $D$ are eigenvalues of the Dirac operator.

One sees that although in the limit $b \rightarrow 0, L \rightarrow
\infty$ the eq.(\ref{1}) is exact, the convergence is not very fast. For
numerical simulations it is important to make the corrections as small
as possible, so one could use a one-dimensional lattice with relatively
small number of
sites $N$. Recently there were several publications, mainly based on the
L\"usher's proposal \cite{ML} for bosonization of fermion
determinants, where different aspects of numerical simulations in the
bosonized models were discussed  \cite{ML1}, \cite{A} \cite{JJL}. Further
development in ref. \cite{M}, \cite{BF}, \cite{BFG} lead to algorithms 
with better convergence properties.

In the present paper I propose a modified algorithm, using essentially
the same idea as in ref. \cite{AS}, but providing much better
convergence. Instead of polynomial supression of finite size effects as
in eq.(\ref{6}), new algorithm provides exponential damping
\begin{equation}
\exp \{-mL \} \label{7}
\end{equation}
where $m$ is a bare quark mass.

\section{Improved algorithm for lattice QCD}

Having in mind applications to QCD in this section we consider two
fermion flavours interacting vectorially with the Yang-Mills fields
$U_{\mu}$. The reason to consider two degenerate flavours is the
positivity of the square of the gauge covariant Dirac operator.
It is convinient to present the square of fermion determinant in the
followi
ng form
 \begin{equation}
\int \exp \{a^4 \sum_{n=1}^2 \sum_x \bar{\psi}_n(x)(
\hat{D}+m) \psi_n(x) \}d \bar{ \psi}d \psi = \det[\gamma_5( \hat{D}+m)]^2=
\label{8} \end{equation}
$$
= \int \exp \{a^4 \sum_x \bar{\psi}(x)(
\hat{D}^2-m^2) \psi(x) \} d \bar{ \psi}d \psi
$$

We again introduce five dimensional bosonic fields $ \phi(x,t)$ with the
same spinorial and internal structure as $ \psi(x)$.  The notations are
as above.

We are going to prove the folowing identity 
\begin{equation} \int \exp \{a^4 \sum_x \bar{ \psi}(x)( \hat{D}^2-m^2)
\psi(x) \}d \bar{\psi}d \psi = \label{9} \end{equation}
$$
= \lim_{L \rightarrow \infty, b \rightarrow 0} \int \exp \{a^4b
\sum_{n=-N+1}^N \sum_x [(b^{-2})( \phi^*_{n+1}(x)  \phi_n(x)+ h.c.-2 \phi^*_n
\phi_n) + $$ $$ -i \phi^*_{n+1}(x) \gamma_5 \hat{D}L^{-1}(2n+1) \phi_n +
h.c. $$
$$
+1/2 \phi^*_n(x) \hat{D}^2L^{-2}(2n+1)^2b^2 \phi_n $$
$$
 +i( \frac{2m}{ \pi L^5})^{1/4}( \phi^*_n(x) \chi(x)+
\chi^*(x) \phi_n(x))2nb \exp \{-mL^{-1}b^2n^2 \}] \}
d \phi^*_nd \phi_nd \chi^*d \chi .  $$ 
Integration goes over the fields $\phi_n(x)$ satisfying the free
boundary conditions (\ref{3}). The last term in the exponent in
the r.h.s. of eq.(\ref{9}) introduces
the constraint
\begin{equation}
\sum_n \phi_nn \exp \{-mL^{-1}b^2n^2 \}=0 \label{10}
\end{equation}
Substituting the solution of this constraint to the eq.(\ref{9}) one
gets the representation for the determinant of the gauge covariant Dirac
operator as the path integral of the exponent of the bosonic Hermitean
action.
Below we present a construction which justifies the eq.(\ref{9}) and
estimate the corrections due to the finite lattice spacing $b$ and
finite lattice size $L$.

Our starting point is the integral
\begin{equation} I= \int \exp \{a^4b
\sum_{n=-N+1}^N \sum_x [(b^{-2})( \phi^*_{n+1}(x) \exp \{-i
\gamma_5 \hat{D}b^2(n+1)^2L^{-1} \} \times \label{11}
\end{equation}
$$ 
 \exp \{i \gamma_5\hat{D}b^2n^2L^{-1} \} \phi_n(x)+ h.c.-2 \phi^*_n
\phi_n) + 
$$
 $$ +i( \frac{2m}{ \pi L^5})^{1/4}( \phi^*_n(x) \chi(x)+
\chi^*(x) \phi_n(x))
2nb \exp \{-mL^{-1}b^2n^2 \}] \}
d \phi^*_nd \phi_nd \chi^*d \chi .  $$ The operator $ \gamma_5 \hat{D}$
is Hermitean and it's eigenvalues are real. R.h.s. of
eq.  (\ref{11}) can be written in terms of eigenvectors of the
operator $ \gamma_5 \hat{D}$, which are denoted as $ \phi_{\alpha}$:
\begin{equation} 
I= \int \exp \{b
\sum_{n=-N+1}^N \sum_{\alpha} [(b^{-2})( \phi^{\alpha*}_{n+1} \exp \{-i
D^{ \alpha}b^2(n+1)^2L^{-1} \} \times \label{12}
\end{equation}
$$ 
 \exp \{iD^{ \alpha}b^2n^2L^{-1} \} \phi_n^{\alpha}+ h.c. -2 \phi^{\alpha*}_n
\phi^{\alpha}_n)+ 
$$
$$
 +i( \frac{2m}{ \pi L^5})^{1/4}( \phi^{\alpha*}_n \chi^{\alpha}+
\chi^{\alpha*} \phi_n^{\alpha})2nb \exp \{-mL^{-1}b^2n^2 \}] \}
d \phi^{\alpha*}_nd \phi^{\alpha}_nd \chi^{\alpha*}d \chi^{\alpha}
$$
To calculate the integral
(\ref{12}) we make the following change of variables:  \begin{equation}
\phi_n^{\alpha} \rightarrow \exp \{-iD^{\alpha}n^2b^2L^{-1} 
\}\phi_n^{\alpha}, \quad \phi_n^{\alpha*} \rightarrow \exp 
\{iD^{\alpha}n^2b^2L^{-1} \} \phi_n^{\alpha*} \label{13} 
\end{equation} Then the integral (\ref{12}) acquires the form 
\begin{equation} I= \int \exp
\{b \sum_{n=-N}^{N} \sum_{\alpha}b^{-2}[ \phi^{* \alpha}_{n+1} 
 \phi^{\alpha}_{n}+ \phi^{* \alpha}_n \phi^{ \alpha}_{n+1} -2 \phi^{* 
 \alpha}_n \phi^{ \alpha}_n] + \label{14} \end{equation} $$ +i( 
\frac{2m}{ \pi L^5})^{1/4}( \phi^{\alpha*}_n \exp 
\{iD^{\alpha}b^2n^2L^{-1} \}\chi^{\alpha}+
\chi^{\alpha*} \exp \{-iD^{\alpha}b^2n^2L^{-1}\phi_n^{\alpha}) \times
$$
$$2nb \exp \{-mL^{-1}b^2n^2 \} \}
d \phi^{\alpha*}_nd \phi^{\alpha}_nd \chi^{\alpha*}d \chi^{\alpha}
$$
Now the quadratic form in the
exponent does not depend on $D^{\alpha}$ and therefore the corresponding
determinant is a trivial constant. So we can calculate the integral by
finding the stationary point of the exponent, which is defined by the
following equations:  $$ b^{-2}( \phi^{* \alpha}_{n+1}+ \phi^{*
\alpha}_{n-1}-2 \phi^{* \alpha}_n)+$$
$$i( \frac{2m}{ \pi L^5})^{1/4} \chi^{*
\alpha}2nb \exp \{-(iD^{\alpha}+m)b^2n^2L^{-1} \}
=0, \quad 
n \neq -N $$ \begin{equation} b^{-2}( \phi^{
\alpha}_{n+1}+ \phi^{ \alpha}_{n-1}-2 \phi^{ \alpha}_n)+ \label{15}
\end{equation}
$$
+i( \frac{2m}{ \pi L^5})^{1/4} \chi^{
\alpha}2nb \exp \{(iD^{\alpha}-m)b^2n^2L^{-1} \}
=0, \quad n \neq -N 
$$
 $$ b^{-2}( \phi^{* \alpha}_{-N+1}-2 \phi^{*
\alpha}_{-N})+i( \frac{2m}{ \pi L^5})^{1/4} \chi^{*
\alpha}2Nb \exp \{-(iD^{\alpha}+m)b^2N^2L^{-1} \}=0
 $$  $$ b^{-2}( \phi^{ \alpha}_{-N+1}-2 \phi^{
 \alpha}_{-N})+i( \frac{2m}{ \pi L^5})^{1/4} \chi^{
\alpha}2Nb \exp \{(iD^{\alpha}-m)b^2N^2L^{-1} \}=0 $$
 $$ \phi_{-N}^{\alpha}= \phi^{ \alpha*}_{-N}=0, \quad \phi^{ \alpha}_{N+1}=
 \phi^{ \alpha*}_{N+1}=0 $$ 

 These equations are most easily solved for small $b$, when they can 
 be approximated by the differential equations:  \begin{equation} 
 \ddot \phi^{* \alpha} +i( \frac{2m}{ \pi L^5})^{1/4} \chi^{* 
\alpha}2t \exp \{-(iD^{\alpha}+m)t^2L^{-1} \}=0\label{16}
  \end{equation} $$ \ddot \phi^{ \alpha} +i( \frac{2m}{ \pi 
L^5})^{1/4} \chi^{ \alpha}2t \exp \{(iD^{\alpha}-m)t^2L^{-1} \}=0 $$ 
$$ \phi^{\alpha}( \frac{L}{2})= \phi^{\alpha}( -\frac{L}{2})=0, \quad 
  \phi^{\alpha*}( \frac{L}{2})= \phi^{\alpha*}(- \frac{L}{2})=0 $$ 
The solution of these eq.s is \begin{equation} \phi^{* \alpha}_{st}= 
i \chi^{* \alpha} \frac{(2m \pi L)^{1/4}}{2(m+iD^{\alpha})^{3/2}} 
\Phi(tL^{-1/2}(m+iD^{\alpha})^{1/2})+ \label{17}
\end{equation}
$$
+ i \chi^{\alpha*} \frac{(2m)^{1/4}}{( \pi L)^{1/4}(m+iD^{\alpha})^2} \exp
\{ \frac{-L(m+iD^{\alpha})}{4} \}2tL^{-1}-i \chi^{\alpha*} \frac{(2 
\pi mL)^{1/4}}{2(m+iD^{\alpha})^{3/2}} $$ \begin{equation} \phi^{ 
\alpha}_{st}= i \chi^{ \alpha} \frac{(2m
\pi L)^{1/4}}{2(m-iD^{\alpha})^{3/2}}
\Phi(tL^{-1/2}(m-iD^{\alpha})^{1/2})+ \label{17a}
\end{equation}
$$
+ i \chi^{\alpha} \frac{(2m)^{1/4}}{( \pi L)^{1/4}(m-iD^{\alpha})^2} \exp
\{ \frac{-L(m-iD^{\alpha})}{4} \}2tL^{-1}-i \chi^{\alpha} \frac{(2 
\pi mL)^{1/4}}{2(m-iD^{\alpha})^{3/2}} $$ where $ \Phi(x)$ is a 
Fresnel integral.  Substituting this solution into the integral 
 (\ref{15}) we get \begin{equation} I= \int \exp\{- \sum_{\alpha} 
 \int_{-L/2}^{+L/2}dt \phi^{\alpha*}_{st}(t)\ddot 
\phi^{\alpha}_{st}(t) \}d \chi^{\alpha*}d \chi^{\alpha} \label{18} 
\end{equation}
Integrating by parts and using the fact that $ 
\phi^{\alpha}_{st}(L/2)= \phi^{\alpha}_{st}(-L/2)=0$, we can 
transform this integral to the form \begin{equation} I= \int \exp \{ 
\sum_{ \alpha} \int_{-L/2}^{L/2}dt \dot \phi^{\alpha*}_{st} \dot 
\phi^{\alpha}_{st} \} d \chi^{\alpha*} d \chi^{\alpha}= \label{19} 
\end{equation}
$$ 
\int \exp \{- \sum_{\alpha} (\frac{2m}{ \pi L})^{1/2} \frac{ \chi^{\alpha*}
\chi^{\alpha}}{m^2+(D^{\alpha})^2}[ \int_{-L/2}^{L/2}dt \exp
\{-2mL^{-1}t^2 \}+ $$
$$
+ \frac{4}{L(m^2+(D^{\alpha})^2)}e^{\{- \frac{Lm}{2} \}}+
 \frac{2 \exp \{- \frac{L(m+iD^{\alpha})}{4}\}}{(m+iD^{\alpha})L}
\int_{-L/2}^{L/2}dt \exp \{-(m-iD^{\alpha})L^{-1}t^2 \}+
$$
$$  \frac{2 \exp \{- \frac{L(m-iD^{\alpha})}{4}\}}{(m-iD^{\alpha})L}
\int_{-L/2}^{L/2}dt \exp \{-(m+iD^{\alpha})L^{-1}t^2 \}] \} 
$$
Integrating over $ \chi$ one gets
\begin{equation}
I= \prod_{\alpha}(m^2+(D^{\alpha})^2)(1+O((mL)^{-3/2} \exp \{-
\frac{mL}{2} \}))  \label{20}
\end{equation}
So if $m \sim a^{-1}$, and $L=2Nb$, the corrections are
of the order
\begin{equation}
O((Nba^{-1})^{-3/2} \exp \{-Nba^{-1} \}) \label{21}
\end{equation}

At the same time it is easy to show that replacing the sum in
the eq.( \ref{15}) by the integral over $t$ produces  corrections of
order $O(b^2a^{-2})$. Therefore taking $N \sim3ab^{-1}$ we can make the
corrections due to the finite size of a lattice less than  1 \%.

Taking into account that 
\begin{equation}
( \gamma_5 \hat{D})^2 =-( \hat{D})^2 \label{23}
\end{equation}
we see that \begin{equation}
I= \det( \hat{D}+m)^2 +O(b^2a^{-2})+O((mL)^{-3/2} \exp \{- 
\frac{mL}{2} \}) \label{23} \end{equation} To conclude we note that 
our equation (\ref{9}) is a linearized version of the integral 
(\ref{11}). The exponent in the quadratic form of the action 
(\ref{11}) may be written in the form \begin{equation} \exp \{-i 
 \gamma_5 \hat{D}b^2L^{-1}(2n+1) \}, \quad 2nb \leq L \label{2 4} 
\end{equation}
 Having in mind that 
\begin{equation}
|| \gamma_5 \hat{D}|| \leq8a^{-1} \label{25}
\end{equation}
one sees that 
\begin{equation}
 \frac{|| \gamma_5 \hat{D}b^2(2n+1)||}{L} \leq 8ba^{-1} \label{26}
\end{equation}
and if $8b<a$ one can replace the exponential in the action (\ref{11})
by the first terms of it's Taylor series, leading to the eq.(\ref{9}).

\section {Discussion}

We proved that the purely bosonic integral (\ref{9}) is equal to the
square of the determinant of covariant Dirac operator up to the terms
\begin{equation}
O(b^2a^{-2})+O((mL)^{- \frac{3}{2}} \exp \{- \frac{mL}{2} \})\label{27}
\end{equation}
Convergence of this approximation is much faster than in the ref.
\cite{AS}, where the finite size effects were supressed polynomially. For 
finite lattice spacing $b$ the corrections decrease exponentially when the 
 number of extra fields $N$ increases.To get the same accuracy one 
needs much smaller number of bosonic fields, which hopefully will 
simplify Monte-Carlo simulations.  It is worthwhile to note that to 
get exponential damping it is crucial to introduce the exponential 
factor into the constraint (\ref{10}). As for the choice of the 
quadratic form in the effective bosonic action, there exists a 
certain freedom. We choose the particular form (\ref{11})to simplify 
analitic calculations.  Presumably the choice of the quadratic form 
done in our previous paper \cite{AS} is also possible. We also note 
that contrary to our previous algorithm \cite{AS} the present 
construction cannot be reduced to L\"usher's algorithm with a 
particular choice of a polynomial due to explicit dependence of the 
constraint on the mass.

{\bf Acknowledgements.} \\
The main part of this work was done while the author was visiting 
Max-Planck Institute for Physics in Munich and Universite Paris-Sud Centre 
d'Orsay. I am grateful to the Theoretical Departments of these Institutes, 
and in particular to D.Maison and Ph.Boucaud for hospitality and 
interesting discussions. I thank A.Galli for helpful comments.This 
researsh was supported in part by Russian Basic Research Fund under 
grant 96-01-00551.$$ ~ $$ \begin{thebibliography}{99} {\small 
\bibitem{AS} A.A.Slavnov SMI-28-95 hep-th/ 9512101, \bibitem{ML} 
M.L\"usher Nucl.Phys.  B418 (1994) 637.  \bibitem{ML1} B.Bunk, 
K.Jansen, B.Jegerlehner, M.L\"usher, H.Simma, R.Sommer, Nucl.Phys.B 
(Proc.Suppl.)42 (1995) 49, \bibitem{A} C.Alexandrou, A.Borelli, Ph. 
de Forcrand, A.Galli and F.Jegerlehner, Nucl.Phys B456 (1995) 296, 
\bibitem{JJL} K.Jansen, B.Jegerlehner, C.Liu hep-lat/9604016, 
\bibitem{M} I.Montvay, DESY 95-192, hep-lat/9510042, \bibitem{BF} 
A.Borici, Ph.de Forcrand IPS-95-23, hep-lat/9509080, \bibitem{BFG} 
A.Borelli, Ph.de Forcrand, A.Gall i, hep-lat/9602016  } \end 
{thebibliography} \end{document}